\documentstyle[twoside,fleqn,espcrc2,epsfig,amssymb]{article}

\newcommand{\myvspace}{\vspace{.5ex}}

\def\rmd{{\rm d}}

\def\rme{{\rm e}}
\def\rmO{{\rm O}}

\def\proof{\noindent{\sl Proof:}\kern0.6em}

\def\frac#1#2{\hbox{$#1\over#2$}}
\def\dual{\mathstrut^*\kern-0.1em}

\def\lvec#1{\setbox0=\hbox{$#1$}
    \setbox1=\hbox{$\scriptstyle\leftarrow$}
    #1\kern-\wd0\smash{
    \raise\ht0\hbox{$\raise1pt\hbox{$\scriptstyle\leftarrow$}$}}
    \kern-\wd1\kern\wd0}
\def\rvec#1{\setbox0=\hbox{$#1$}
    \setbox1=\hbox{$\scriptstyle\rightarrow$}
    #1\kern-\wd0\smash{
    \raise\ht0\hbox{$\raise1pt\hbox{$\scriptstyle\rightarrow$}$}}
    \kern-\wd1\kern\wd0}

\def\nab#1{{\nabla_{#1}}}
\def\nabstar#1{\nabla\kern-0.5pt\smash{\raise 4.5pt\hbox{$\ast$}}
               \kern-4.5pt_{#1}}
\def\drv#1{{\partial_{#1}}}
\def\drvstar#1{\partial\kern-0.5pt\smash{\raise 4.5pt\hbox{$\ast$}}
               \kern-5.0pt_{#1}}

\def\MeV{{\rm MeV}}

\def\fm{{\rm fm}}

\def\Nf{N_{\rm f}}
\def\psibar{\overline{\psi}}

\def\rhoprime{\rho\kern1pt'}
\def\rhobar{\bar{\rho}}
\def\rhobarprime{\rhobar\kern1pt'}
\def\rhobartilde{\kern2pt\tilde{\kern-2pt\rhobar}}
\def\rhobartildeprime{\kern2pt\tilde{\kern-2pt\rhobar}\kern1pt'}

\def\zetabar{\bar{\zeta}}
\def\zetaprime{\zeta\kern1pt'}
\def\zetabarprime{\zetabar\kern1pt'}
\def\zetar{\zeta_{\raise-1pt\hbox{\sixrm R}}}
\def\zetabarr{\zetabar_{\raise-1pt\hbox{\sixrm R}}}
\def\phieff{\phi_{\rm eff}}
\def\phiimpr{\phi_{\kern0.5pt\hbox{\sixrm I}}}

\def\ar{A_{\mbox{\scriptsize{\rm R}}}}
\def\vr{V_{\mbox{\scriptsize{\rm R}}}}
\def\pr{P_{\mbox{\scriptsize{\rm R}}}}

\def\ai{A_{\mbox{\scriptsize{\rm I}}}}

\def\dirac#1{\gamma_{#1}}
\def\diracstar#1#2{
    \setbox0=\hbox{$\gamma$}\setbox1=\hbox{$\gamma_{#1}$}
    \gamma_{#1}\kern-\wd1\kern\wd0
    \smash{\raise4.5pt\hbox{$\scriptstyle#2$}}}

\def\ba{b_{\rm A}}
\def\bv{b_{\rm V}}
\def\bp{b_{\rm P}}

\def\ca{c_{\rm A}}
\def\cv{c_{\rm V}}
\def\csw{c_{\rm sw}}

\def\fa{f_{\rm A}}

\def\fIaa{f_{\rm AA}^{\rm I}}

\def\fp{f_{\rm P}}

\def\fIv{f_{\rm V}^{\rm I}}
\def\f1{f_1}
\def\kv{k_{\rm V}}
\def\kt{k_{\rm T}}

\def\kaaI{k_{\rm AA}^{\rm I}}

\def\tr{\,\hbox{tr}\,}

\def\CF{C_{\rm F}}

\def\Sf{S_{\rm F}}
\def\Seff{S_{\rm eff}}

\def\op#1{{\cal O}_{\rm #1}}
\def\opprime#1{\setbox0=\hbox{${\cal O}$}\setbox1=\hbox{${\cal O}_{\rm #1}$}
    {\cal O}_{\rm #1}\kern-\wd1\kern\wd0
    \smash{\raise4.5pt\hbox{\kern1pt$\scriptstyle\prime$}}\kern1pt}

\def\ophatprime#1{\setbox0=\hbox{$\widehat{\cal O}$}
    \setbox1=\hbox{$\widehat{\cal O}_{\rm #1}$}
    \widehat{\cal O}_{\rm #1}\kern-\wd1\kern\wd0
    \smash{\raise4.5pt\hbox{\kern1pt$\scriptstyle\prime$}}\kern1pt}

\def\bopprime#1{\setbox0=\hbox{${\cal O}$}\setbox1=\hbox{${\cal O}_{\rm #1}$}
    {\cal L}_{\rm #1}\kern-\wd1\kern\wd0
    \smash{\raise4.5pt\hbox{\kern1pt$\scriptstyle\prime$}}\kern1pt}

\def\blagprime#1{\setbox0=\hbox{${\cal B}$}\setbox1=\hbox{${\cal B}_{#1}$}
    {\cal B}_{#1}\kern-\wd1\kern\wd0
    \smash{\raise5.2pt\hbox{\kern1pt$\scriptstyle\prime$}}\kern1pt}

\def\gbar{\bar{g}}

\def\mq{m_{\rm q}}

\def\mr{m_{{\mbox{\scriptsize{\rm R}}}}}
\def\mc{m_{\rm c}}

\def\m{m}
\def\mprime{m'}

\def\r{r}
\def\s{s}
\def\rprime{r'}
\def\sprime{s'}
\def\M{M}
\def\Mprime{M'}
\def\dM{\Delta M}

\def\za{Z_{\rm A}}
\def\zp{Z_{\rm P}}
\def\zv{Z_{\rm V}}

\def\zm{Z_{\rm m}}

\def\zphi{Z_{\phi}}

\def\gparisi{g_{\rm P}}

\def\msbar{{\rm \overline{MS\kern-0.05em}\kern0.05em}}

\def\kc{\kappa_{\rm c}}
\def\hopc{\kappa_{\rm c}}
\def\hop{\kappa}

\def\Oa{\rmO(a)}
\def\nf{N_{\rm f}}

\title{%
\vspace{-3.1cm}
\begin{flushleft}
       {\normalsize CERN--TH/97--107}  \\[-0.2cm]
       {\normalsize May 1997}   \\
\end{flushleft}
       \vspace{0.7cm}
O($a$) improved lattice QCD
\thanks{Invited talk presented at the workshop {\sl
   Lattice QCD on parallel Computers}, CCP, Tsukuba, Japan, March 1997.}
}

\author{Rainer Sommer\address{CERN, Theory Division, CH-1211 Gen\`eve 23,
                      Switzerland \\ and 
                      DESY-IfH Zeuthen, Platanenallee 6,
                       D-15738 Zeuthen, Germany}
        }
\begin{document}

\newcommand{\ewxy}[2]{\setlength{\epsfxsize}{#2}\epsfbox[30 30 640 640]{#1}}

\begin{abstract}
  We review the O($a$) improvement of lattice QCD
  with special emphasis on the motivation for performing the
  improvement programme non-perturbatively and the general concepts
  of on-shell improvement.
  The present status of the calculations of various improvement 
  coefficients (perturbative and non-perturbative) is reviewed, as
  well as the computation of the
  isospin current normalization constants $\za$
  and $\zv$.  
  We comment on recent results for hadronic observables 
  obtained in the improved theory.
%%   and on progress in the computation of the
%%  improved action of full QCD.
\end{abstract}
\maketitle

\section{INTRODUCTION}

The leading cutoff effects in lattice QCD with Wilson quarks
\cite{Wilson} are proportional to the lattice spacing $a$ and can be
rather large for typical values of the Monte Carlo (MC) simulation
parameters~\cite{letter,mrhosigma}.  Decreasing the lattice spacing at constant
physical length scales means larger lattices and therefore rapidly
increasing costs of the MC simulations.

An alternative method to reduce cutoff effects in lattice field
theories is due to Symanzik~\cite{SymanzikI,SymanzikII}.  He has shown
that the approach of Green's functions to their continuum limit can be
accelerated by using an improved action and improved (composite)
fields.  A considerable simplification is achieved if the improved
continuum approach is only required for on-shell quantities such as
particle masses 
and matrix elements of improved fields between physical 
states~\cite{OnShell,SW}.

Symanzik improvement may be viewed as an extension of the
renormalization programme to the level of irrelevant operators. While
the structure of the possible counterterms is dictated by the
symmetries, their coefficients have to be fixed by appropriate
improvement conditions. Although they can be estimated in 
(tadpole-improved~\cite{lepenzie_92}) perturbation
theory, a non-perturbative determination of the improvement
coefficients through MC simulations is clearly preferable.

To O($a$), we have recently carried out the non-perturbative
improvement programme in quenched lattice QCD, thus leaving residual
cutoff effects of O($a^2$) only.  The details of the calculations
are given
elsewhere~\cite{paperI}--\cite{paperV}. 
In this report we emphasize the general concepts and give a short
account of the main results. We also discuss progress that is currently made
in applying the improvement programme to full QCD with $\nf=2$
flavours~\cite{csw_dyn}.

We start by outlining the importance of removing the
linear (in $a$) lattice artefacts non-perturbatively (sect.\,2). 
We then review on-shell improvement in the framework of Symanzik's
local effective theory (sect.\,3) and the general concept of
non-perturbative improvement (sect.\,4).  
The practical calculations are done in finite space-time volume with
Schr\"odinger functional boundary conditions which we introduce in
sect.\,5.  We then review the computations of the various improvement 
coefficients
and the current normalization constants in quenched lattice QCD. 
Residual lattice artefacts are 
discussed in sect.\,9 and the status of $\Oa$ improvement in full QCD in
sect.~10.

\section{WHY {\it NON-PERTURBATIVE} $\Oa$ IMPROVEMENT?}
 
In this workshop two main alternative approaches to reducing 
the cutoff effects of lattice QCD are 
discussed~\cite{Nieder_tsuk,Lepage_tsuk}. 
Before giving some detailed motivation for what is discussed here,
let us point out the relation to
these talks.
The goal of the perfect action approach~\cite{Nieder_tsuk} is the 
improvement of the lattice 
theory to all orders of the lattice spacing.
In determining the perfect action, one -- in practice -- performs an expansion 
in powers of the
coupling constant. In contrast, the non-perturbative Symanzik improvement
as discussed here 
proceeds non-perturbatively in the coupling and order by order in the lattice
spacing. On the other hand, ``mean field improvement''~\cite{Lepage_tsuk} 
is Symanzik improvement with improvement coefficients approximated by
mean field-improved perturbation theory~\cite{lepenzie_92}.

Let us explain in more detail the motivation for non-perturbatively
removing the lattice
artefacts that are linear in $a$.  The first motivation 
is that these are large numerically. We give 
two examples to 
illustrate this point.

\begin{figure}[tb]
\vspace{-65pt}
\epsfig{file=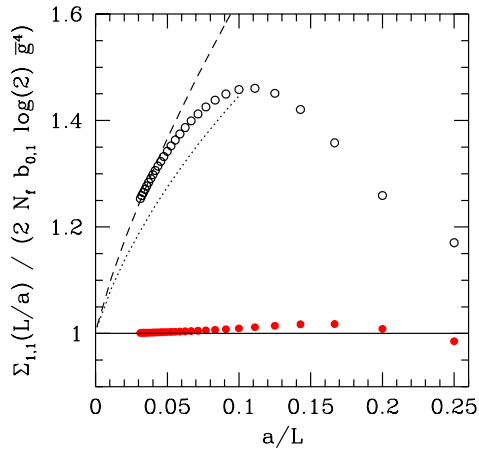,%18 144 592 718
       bbllx=68pt,%
       bblly=144pt,%
       bburx=642pt,%
       bbury=768pt,%
        clip=,%
       width=9cm}%
%      height=10cm}
%\framebox[55mm]{\rule[-21mm]{0mm}{43mm}}
\vspace{-60pt}
\caption{One-loop contribution of a massless Wilson quark to the step scaling 
  function $\Sigma$ normalized to its continuum value.
  Open circles are for the Wilson action. 
  The dotted curve describes the leading lattice artefact 
  for small $a/L$, which is of the form $\log(a/L) a/L $, while the
  dashed curve contains the pure $a/L$-term in addition.
  Filled circles illustrate the behaviour after $\Oa$-improvement.}
\label{f_arte}
\end{figure}

The first example is a perturbative one. 
Consider the renormalized coupling $\gbar(L)$ defined in the Schr\"odinger 
functional 
 scheme~\cite{alphaI,StefanI,StefanRainer} for massless
fermions. Its evolution from length scale $L$ to
scale $2L$ defines
the step scaling function $\Sigma$~\cite{alphaI,alphaII}. 
The one-loop contribution of $\Nf$
Wilson fermions to $\Sigma$ has been calculated in ref.~\cite{StefanRainer}.
We denote it by $\gbar^4 \Nf \Sigma_{1,1}$~. Its continuum limit is given by the
fermion contribution to the lowest term in the $\beta$-function:
\begin{eqnarray}
 \lim_{a/L\rightarrow 0} {\Sigma_{1,1}(L/a) \over \gbar^4 \Nf } = 2 b_{0,1} 
 \log(2)
 \, , \quad \\
 b_{0,1} = -{1 \over 24\pi^2}\, . \nonumber 
\end{eqnarray}
Without $\Oa$-improvement, $\Sigma_{1,1}$ depends very strongly on $a/L$ and
shows a smooth approach to the continuum only for very large lattices 
(cf. fig.~\ref{f_arte}). 
If one had MC data with $L/a\leq10$ and such an error term, 
a naive continuum extrapolation of such data
would overestimate the fermion
contribution by O(60\%). On the other hand, after $\Oa$-improvement the 
lattice artefacts in this quantity are tiny. 

\begin{figure}[tb]
\vspace{5pt}
\epsfig{file=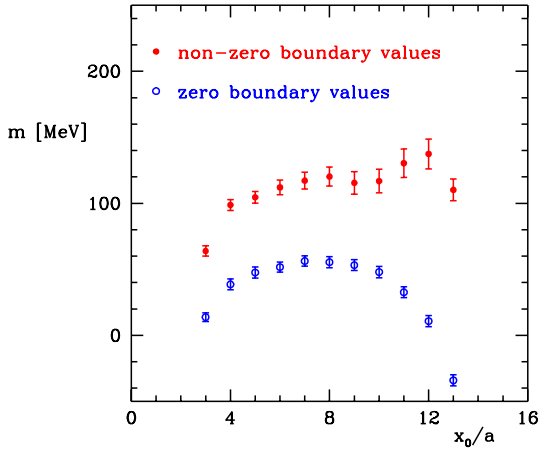,%18 144 592 718
       width=7.0cm} % trim it a little; normal:7.5
%      height=10cm}
\vspace{-30pt}
\caption{Dependence of curent quark mass $m$ on the boundary 
  condition and the time coordinate. The calculation is done 
  on a $16 \times 8$ lattice at $\beta=6.4$, which corresponds to a
  lattice spacing of $a \approx 0.05\,\fm$. 
  ``Boundary values'' refer to the 
  gauge field boundary conditions in the SF~\cite{letter};
  see sects.~\ref{s_SF},\ref{s_IAAC} for details.}
\label{f_letter}
\end{figure}

The second example is the current quark mass $m$
defined by the PCAC relation. As we will discuss in
detail below, its value is independent of kinematical variables
such as the boundary conditions. Dependences on such 
variables are pure lattice artefacts. 
We examined the current quark mass in the valence
approximation by numerical Monte Carlo simulations and found~\cite{letter}
large lattice artefacts even for quite small lattice spacings
(cf. fig.~\ref{f_letter}).

Since lattice artefacts can be large numerically, it is desirable to remove them
non-perturbatively in order not to leave over $\alpha_s^2 a$ 
or truly non-perturbative terms that may be noticeable.

A second motivation for performing improvement non-perturbatively is closely
related. After improvement one wants to compute physical quantities $P$, such as 
ratios of hadron masses for a range of lattice spacings and extrapolate them to
the continuum limit. 

For example, assume that MC results $P(a)$ are available
for four points $a^2 = a_0^2, \frac34 a_0^2, \frac12 a_0^2, \frac14 a_0^2$ and
with statistical errors $\Delta P = \Delta$ that are independent of $a$.
An extrapolation by means of a fit $P(a)= P_{\rm c} + s_2 a^2 $ will yield an 
estimate
for the continuum limit of $P$ with error $\Delta P_{\rm c} = 1.22\Delta$. 
However, if the calculation was performed with approximate values of the
improvement coefficients, a small (?) linear error term $b \times a $ will be
left in the results $P(a)$. If ignored, it causes a systematic error of size
$0.35\, b\, a_0$
in $P_{\rm c}$. As an alternative, one may, of course, perform a 
fit with  the ansatz $P(a)= P_{\rm c} + s_1 a  + s_2 a^2$. This gives a 
continuum value
with statistical error $\Delta P_{\rm c} = 9.7\Delta$.

Trivial as it is, the above example illustrates the important point:
in order to perform a reliable continuum extrapolation, one must have theoretical
information on the expected $a$-dependence. This is contained 
in the structure of Symanzik's effective action, which 
justifies a power series in the lattice spacing for the correction terms.
(For the purpose of extrapolating MC-data, the logarithmic dependence 
on $a$ can be ignored).
Under the same assumption, namely that lattice artefacts higher than $a^2$ can
be neglected, the continuum extrapolation in the $\Oa$-improved theory 
gives errors that are smaller by a factor 8 than the extrapolation in
the unimproved (or approximately improved) case.

Natural as it may seem to continue the non-perturbative Symanzik improvement
beyond the first order in the lattice spacing, this 
appears impossible in practice. The reason is that 
at order O($a^2$) a large number of improvement terms would
have to be determined by MC calculations. 

\section{ON-SHELL IMPROVEMENT \label{s_OSI}}
\subsection{Lattice QCD with Wilson quarks}

In this section we consider QCD on an infinitely extended lattice with
two degenerate light Wilson quarks of bare mass $m_0$~\cite{Wilson}.
The action is the sum of the usual Wilson plaquette action and the
quark action
\begin{equation}
 \Sf[U,\psibar,\psi\,]=a^4\sum_{x}\psibar(x)(D+m_0)\psi(x),
 \label{e_quark}
\end{equation}
where $a$ denotes the lattice spacing.  The Wilson-Dirac operator
\begin{equation}
  D=\frac12\sum_{\mu=0}^3\bigl[(\nabstar\mu
  +\nab\mu)\dirac\mu-a\nabstar\mu\nab\mu\bigr],
\label{e_Wilson-Dirac}
\end{equation}
contains the lattice covariant forward and backward derivatives,
$\nab\mu$ and $\nabstar\mu$.  The last term in
eq.\,(\ref{e_Wilson-Dirac}) eliminates the unwanted doubler states but
also breaks chiral symmetry.  As a consequence, both additive and
multiplicative renormalization of the quark mass are necessary,
i.e.~any renormalized quark mass $\mr$ is of the form
\begin{equation}
 \mr=\zm \mq,\qquad \mq=m_0-\mc,
\end{equation}
where $\mc$ is the so-called critical quark mass. 

Chiral symmetry violation is more directly seen by studying the
conservation of the isovector axial current $A_\mu^a$. The current and
the associated axial density on the lattice are defined through
\begin{eqnarray}
  A_\mu^a(x)&=&\psibar(x)\dirac\mu\dirac5{{\tau^a}\over{2}}\psi(x),
  \label{e_axialcurrent}\\
  P^a(x)&=&\psibar(x)\dirac5{{\tau^a}\over{2}}\psi(x),
\end{eqnarray}
where $\tau^a$ are Pauli matrices
acting on the flavour index of the quark field. The PCAC relation
\begin{eqnarray}
  \tilde{\partial}_\mu A_\mu^a(x) &=& 2mP^a(x) + \rmO(a), \label{e_PCAC_wils}\\
  \tilde{\partial}_\mu &=& \frac{1}{2}(\drvstar{\mu}+\drv{\mu})
  \label{e_PCAC} 
\end{eqnarray}
then includes an error term of order~$a$, which can be rather large on
the accessible lattices\,\cite{letter}, as shown in fig.~\ref{f_letter}.
Above, $\drv{\mu}$ and $\drvstar{\mu}$ denote forward and backward lattice 
derivatives and $m$ is the unrenormalized current quark mass at scale $1/a$.

The isospin symmetry remains unbroken on the lattice and there exists
an associated conserved vector current. However, it is often
advantageous to use the current which is strictly local,
\begin{equation}
 V_\mu^a(x)=\psibar(x)\dirac\mu{{\tau^a}\over{2}}\psi(x).
\end{equation}
The conservation of this current is then also violated by cutoff
effects, and a finite renormalization is required to ensure that the
associated charge takes half-integral values.
	
\subsection{Symanzik's local effective theory}

Near the continuum limit the lattice theory can be described in terms
of a local effective theory~\cite{SymanzikII},
\begin{equation}
 \Seff=S_0+a S_1+a^2 S_2+\ldots,
 \label{e_Seff}
\end{equation}
where $S_0$ is the action of the continuum theory, defined e.g.~on a
lattice with spacing $\varepsilon\ll a$.  The terms $S_k$,
$k=1,2,\ldots$, are space-time integrals of Lagrangians ${\cal
  L}_k(x)$.  These are given as general linear combinations of local
gauge-invariant composite fields which respect the exact symmetries of
the lattice theory and have canonical dimension $4+k$.  We use the
convention that explicit (non-negative) powers of the quark mass $m$
are included in the dimension counting.  A possible basis of fields
for the lagrangian ${\cal L}_1(x)$ then reads
\begin{eqnarray}
{\cal O}_1 &=&\psibar\,\sigma_{\mu\nu}F_{\mu\nu}\psi,\nonumber\\[.2ex]
{\cal O}_2 &=&\psibar\,D_{\mu}D_{\mu}\psi
           +\psibar\,\lvec{D}_{\mu}\lvec{D}_{\mu}\psi,\nonumber\\[.2ex]
{\cal O}_3 &=& m\tr\!\left\{F_{\mu\nu}F_{\mu\nu}\right\},
              \label{e_counterterms}\\[.2ex]
{\cal O}_4 &=& m\left\{\psibar\,\dirac{\mu}D_{\mu}\psi
               -\psibar\,\lvec{D}_{\mu}\dirac{\mu}\psi\right\},
              \nonumber\\[.2ex]
{\cal O}_5 &=& m^2\psibar\psi,\nonumber
\end{eqnarray} 
where $F_{\mu\nu}$ is the field tensor and
$\sigma_{\mu\nu}=\frac{i}2[\dirac\mu,\dirac\nu]$.

When considering correlation functions of local gauge invariant fields
the action is not the only source of cutoff effects.  If $\phi(x)$
denotes such a lattice field (e.g.~the axial density or the isospin
currents of subsect.\,\ref{s_OSI}.1), one expects the connected $n$-point
function
\begin{equation}
  G_n(x_1,\ldots,x_n)=(\zphi)^n
  \left\langle\phi(x_1)\ldots\phi(x_n)\right\rangle_{\rm con}
\end{equation}
to have a well-defined continuum limit, provided the renormalization
constant $\zphi$ is correctly tuned and the space-time arguments
$x_1,\ldots,x_n$ are kept at a physical distance from each other.

In the effective theory the renormalized lattice field $\zphi\phi(x)$
is represented by an effective field,
\begin{equation}
  \phieff(x)=\phi_0(x)+a\phi_1(x)+a^2\phi_2(x)+\ldots,
\end{equation} 
where the $\phi_k(x)$ are linear combinations of composite, local
fields with the appropriate dimension and symmetries.  For example, in
the case of the axial current~(\ref{e_axialcurrent}), $\phi_1$ is
given as a linear combination of the terms
\begin{eqnarray}
  (\op{6})_{\mu}^a &=&
  \psibar\,\dirac{5}\frac{1}{2}\tau^a\sigma_{\mu\nu}
      \bigl[{D}_{\nu}-\lvec{D}_{\nu}\bigr]\psi,
  \nonumber\\[.5ex]
  (\op{7})_{\mu}^a&=&\psibar\,\frac{1}{2}\tau^a\dirac{5}
                  \bigl[{D}_{\mu}+\lvec{D}_{\mu}\bigr]\psi,
  \label{e_impr_current}\\[.5ex]
  (\op{8})_{\mu}^a&=&m\psibar\,\dirac{\mu}\dirac{5}\frac{1}{2}\tau^a\psi.
  \nonumber
\end{eqnarray}
The convergence of $G_n(x_1,\ldots,x_n)$
to its continuum limit can now be studied in the 
effective theory,
\begin{eqnarray}
  \lefteqn{G_n(x_1,\ldots,x_n) 
  =\left\langle\phi_0(x_1)\ldots\phi_0(x_n)\right\rangle_{\rm con}}
  \nonumber\\[.3ex]
  &&\mbox{}-a\int\rmd^4y\,\left\langle\phi_0(x_1)\ldots\phi_0(x_n)
  {\cal L}_1(y)\right\rangle_{\rm con}
  \nonumber\\
  &&\mbox{}+a\sum_{k=1}^n
  \left\langle\phi_0(x_1)\ldots\phi_1(x_k)\ldots\phi_0(x_n)
  \right\rangle_{\rm con}
  \nonumber\\
  &&\hphantom{0123456}+\rmO(a^2),
 \label{e_continuum_approach}
\end{eqnarray}
where the expectation values on the right-hand side 
are to be taken in the continuum theory with action $S_0$.

\subsection{Using the field equations}

For most applications, it is sufficient to
compute on-shell quantities such as particle
masses, S-matrix elements and correlation functions
at space-time arguments, which are separated by
a physical distance. It is then possible to make use of the
field equations to reduce first the number of 
basis fields in the effective Lagrangian ${\cal L}_1$
and, in a second step, also in the O($a$) counterterm
$\phi_1$ of the effective composite fields.

If one uses the field equations in the Lagrangian ${\cal L}_1$ under
the space-time integral in eq.\,(\ref{e_continuum_approach}), the
errors made are contact terms that arise when $y$ comes close to one
of the arguments $x_1,\ldots,x_n$.  Taking into account the dimensions
and symmetries, one easily verifies that these contact terms must have
the same structure as the insertions of $\phi_1$ in the last term of
eq.\,(\ref{e_continuum_approach}). Using the field equations in 
${\cal L}_1$ therefore just means a redefinition of the coefficients
in the counterterm $\phi_1$.

It turns out that one may eliminate two of the terms in
eq.\,(\ref{e_counterterms}). A possible choice is to stay with the
terms ${\cal O}_1$, ${\cal O}_3$ and ${\cal O}_5$, which yields the
effective continuum action for on-shell quantities to order $a$.
Having made this choice one may apply the field equations once again
to simplify the term $\phi_1$ in the effective field as well.  In the
example of the axial current it is then possible to eliminate the
term~$\op{6}$ in eq.\,(\ref{e_impr_current}).

\subsection{Improved lattice action and fields}

The on-shell O($a$) improved lattice action is obtained by
adding a counterterm to the unimproved lattice action
such that the action $S_1$ in the effective
theory is cancelled in on-shell amplitudes. 
This can be achieved by adding lattice representatives
of the terms ${\cal O}_1$, ${\cal O}_3$ and ${\cal O}_5$
to the unimproved lattice Lagrangian, with coefficients that
are functions of the bare coupling $g_0$ only.
Leaving the discussion of suitable improvement conditions to
sect.\,4, we here note that the fields 
${\cal O}_3$ and ${\cal O}_5$ already appear in the unimproved
theory and thus merely lead to a reparametrization of 
the bare parameters $g_0$ and $m_0$. In the following, 
we will not consider these terms any further. 
There relevance in connection with
massless renormalization
schemes is discussed in detail in ref.~\cite{paperI}.

We choose the standard discretization $\widehat{F}_{\mu\nu}$
of the field tensor\,\cite{paperI} and add the improvement term
to the Wilson-Dirac operator~(\ref{e_Wilson-Dirac}),
\begin{equation}
 D_{\rm impr}=D+\csw\,{{ia}\over{4}}\sigma_{\mu\nu}\widehat{F}_{\mu\nu}.
 \label{e_dimpr}
\end{equation} 
With a properly chosen coefficient $\csw(g_0)$, 
this yields the on-shell O($a$) improved lattice action which 
has first been proposed by Sheikholeslami and Wohlert~\cite{SW}.

The perturbative expansion of $\csw$ reads  
$\csw=1+\csw^{(1)}g_0^2+\rmO(g_0^4)$, with~\cite{paperII,Wohlert}
$ \csw^{(1)} = 0.26590(7)$. 

The O($a$) improved isospin currents and the axial density can
be parametrized as follows,
\begin{eqnarray}
 (\ar)_\mu^a\!\!\!&=&\!\!\!
 \za(1+\ba a\mq)\bigl\{A_\mu^a+a\ca\tilde{\partial}_\mu
       P^a\bigr\},\nonumber\\[.5ex]
 (\vr)_\mu^a\!\!\!&=&\!\!\!
 \zv(1+\bv a\mq)\bigl\{V_\mu^a+a\cv\tilde{\partial}_\nu
       T_{\mu\nu}^a\bigr\},\nonumber\\
  [.5ex]
 (\pr)^a\!\!\!&=&\!\!\!
 \zp(1+\bp a\mq)P^a\,, \label{e_renfields}
\end{eqnarray}
where
$$
T^a_{\mu\nu}=i\psibar\sigma_{\mu\nu}\frac12\tau^a\psi
$$
and the other fields
 have been defined in
subsect.\,2.1.
We have
included the normalization constants $Z_{\rm A,V,P}$, which have to be
fixed by appropriate normalization conditions (cf.~sect.\,\ref{s_IAAC}).  
Again,
the improvement coefficients $b_{\rm A,V,P}$ and $c_{\rm A,V}$ are
functions of $g_0$ only. At tree level of perturbation theory, they are
given by
$\ba=\bp=\bv=1$ and $\ca=\cv=0$~\cite{HeatlieEtAl,paperII}. 
Due to the efforts of L\"uscher, Sint and Weisz~\cite{paperII,paperV} these
coefficients are now
known to one-loop accuracy, 
\begin{eqnarray}
\ca &=&-0.005680(2)\times g_0^2\CF+\rmO(g_0^4),\quad \nonumber\\
\cv &=&-0.01225(1)\times g_0^2\CF+\rmO(g_0^4),\quad \nonumber\\
\ba &=& 1 + 0.11414(4)\times g_0^2\CF+\rmO(g_0^4),\quad \nonumber\\
\bv &=& 1 + 0.11492(4)\times g_0^2\CF+\rmO(g_0^4),\quad \nonumber\\
\bp &=& 1 + 0.11484(2)\times g_0^2\CF+\rmO(g_0^4).\quad 
 \label{e_cApert}
\end{eqnarray}
Here $\CF=4/3$. It is not known why $\ba,\bv$ and $\bp$ are numerically so close
to each other to this order of perturbation theory. 

Non-perturbative
determinations of $\csw,\ca,\cv$ and $\bv$ will be discussed in sections~6--8.

\subsection{The PCAC relation}

We assume for the moment that on-shell O($a$) improvement has been
fully implemented, i.e.~the improvement coefficients are assigned
their correct values. If ${\cal O}$ denotes a renormalized on-shell
O($a$) improved field localized in a region not containing $x$, we
thus expect that the PCAC relation
%(with $\tilde{\partial}_\mu=\frac12(\drv\mu+\drvstar\mu)$),
%
\begin{equation}
%    \frac12(\drv\mu+\drvstar\mu)
    \tilde{\partial}_\mu
    \langle(\ar)_\mu^a(x)\,{\cal O}\rangle=
     2\mr\langle(\pr)^a(x)\,{\cal O}\rangle
 \label{e_PCAC_impr}
\end{equation}
holds up to corrections of order $a^2$. 
At this point we note that the field ${\cal O}$ need not
be improved for this statement to be true. To see this
we use again Symanzik's local effective theory and
denote the O($a$) correction term
in ${\cal O}_{\rm eff}$ by $\phi_1$.
Eq.\,(\ref{e_PCAC_impr}) then receives an order $a$ contribution
\begin{equation} 
   a\bigl\langle
  \left\{\partial_\mu(\ar)^a_{\mu}(x)-2\mr(\pr)^a(x)\right\}
  \phi_1\bigr\rangle,
 \label{e_PCACcorrection}
\end{equation}
which is to be evaluated in the continuum theory.
The PCAC relation holds exactly in this limit and
the extra term~(\ref{e_PCACcorrection}) thus
vanishes. 
% a conclusion that will remain true
%in the set-up with the QCD Schr\"odinger functional 
%introduced below.

\section{NON-PERTURBATIVE IMPROVEMENT \label{s_NPI}}

We have explained above how the form of the improved action and the
composite fields is determined by the symmetries
of the lattice action. 
The coefficients of the different terms need to be fixed by suitable
improvement conditions. One considers pure lattice artefacts,
i.e. combinations
of observables that are known to vanish in the continuum limit
of the theory. Improvement conditions require
these lattice artefacts to vanish,
thus defining the values of the improvement 
coefficients as a function of the lattice spacing.

In perturbation theory, lattice artefacts 
can be obtained from any (renormalized) quantity by subtracting its
value in the continuum limit. The improvement coefficients are 
unique.\footnote{One might think there is a scheme ambiguity as
in the general renormalization of a theory. The ``scheme'' has, however, 
already been 
chosen implicitly, when one writes down the original (unimproved) lattice
theory. It changes, for instance, when a different discretization of the pure gauge action
is chosen.} 

Beyond perturbation theory, one wants to determine the improvement coefficients by
MC calculations and it requires significant effort
to take the continuum limit.
It is therefore advantageous to use lattice artefacts that derive from
a symmetry of the continuum field theory that is not respected by the
lattice regularization. One may require rotational invariance
of the static potential $V({\bf r})$, e.g.
$$
 V({\bf r}=(2,2,1)r/3) - V({\bf r}=(r,0,0)) =0 \, ,
$$
or 
Lorentz invariance,
$$
 [E({\bf p})]^2 - [E({\bf 0})]^2  - {\bf p}^2 =0 \, ,
$$
for the momentum dependence of a one-particle energy $E$.

For $\Oa$ improvement of QCD it is advantageous
to use violations of the PCAC relation~(\ref{e_PCAC_wils}), instead.
PCAC can be used in the context of the 
Schr\"odinger functional (SF), where one has a large flexibility to
choose appropriate improvement conditions and can compute the improvement
coefficients also for small values of the bare coupling $g_0$, making contact
with their perturbative expansions. A further -- and maybe 
the most significant --  advantage of the SF in
this context is the following. In the SF we may choose 
boundary conditions 
such that the classical solution, called the induced
background field, has non-vanishing components $F_{\mu\nu}$~\cite{alphaI}. 
Remembering eq.~(\ref{e_dimpr}), we observe that 
correlation functions are then sensitive to the improvement coefficient 
$\csw$ already
at tree level of perturbation theory. In general this will be the case 
at higher orders only. This is the basis for a good sensitivity of
the improvement conditions to $\csw$. 
 
As a consequence of the freedom to choose improvement conditions, 
the resulting values of 
improvement coefficients such as $\csw,\,\ca$
depend on the exact choices made for the
improvement conditions. The corresponding variation of
$\csw,\,\ca$ is of order $a$. This variation changes the effects of
order $a^2$ in physical observables computed after improvement.  In
principle this is irrelevant at the level of $\rmO(a)$ improvement.
Nevertheless, one ought to choose improvement conditions where such
terms have small coefficients. The improvement conditions derived
from the SF can be studied in perturbation theory. Such a study
provided essential criteria for our detailed choice of improvement
conditions. 

In the following section we introduce the SF to prepare 
for the improvement conditions that are detailed in
sects.~\ref{s_IAAC},\ref{s_IVC}.

\section{THE SCHR\"ODINGER FUNCTIONAL \label{s_SF}}

\subsection{Definitions}

The space-time lattice is now taken to be a discretized hyper-cylinder
of length $T$ and circumference $L$.  In the spatial directions the
quantum fields are $L$-periodic, whereas in the Euclidean time
direction inhomogeneous Dirichlet boundary conditions are imposed as
follows.  The spatial components of the gauge field are required to
satisfy
\begin{equation}
  \left.U(x,k)\right|_{x_0=0}=\exp(aC_k), 
 \quad C_k=i\phi/L,
 \label{e_CCprime}
\end{equation}
with $\phi={\rm diag}(\phi_1,\phi_2,\phi_3)$, and an analogous
boundary condition with $C'$ is imposed at $x_0=T$.

With the projectors $P_\pm=\frac12(1\pm\dirac0)$, the boundary
conditions for the quark and antiquark fields read
\begin{eqnarray}
  P_{+}\psi|_{x_0=0}=\rho,
  &&
  P_{-}\psi|_{x_0=T}=\rhoprime,\\[1ex]
  \psibar P_{-}|_{x_0=0}=\rhobar,
  &&
  \psibar P_{+}|_{x_0=T}=\rhobarprime.
\end{eqnarray}
The functional integral in this situation~\cite{alphaI,StefanI},
\begin{equation} 
  {\cal Z}[C',\rhobarprime,\rhoprime;C,\bar\rho,\rho]=
  \int_{\rm fields}\,\rme^{-S},
\end{equation}
is known as the QCD Schr\"odinger functional~(SF).  Concerning the
(unimproved) action $S$, we note that its gauge field part has the
same form as in infinite volume.  The quark action is again given by
eq.\,(\ref{e_quark}), provided one formally extends the
quark and antiquark fields to Euclidean times $x_0<0$ and $x_0>T$ by
``padding" with zeros~\cite{paperI}. However, we will use a slightly
more general covariant lattice derivative,
\begin{equation}
   \nab{\mu}\psi(x)=
  {1\over a}\bigl[\lambda_{\mu}U(x,\mu)\psi(x+a\hat{\mu})-\psi(x)\bigr],
\end{equation}
where $\lambda_0=1$ and $\lambda_k=\exp(ia\theta/L)$.  This
modification of the covariant derivative is equivalent to demanding
spatial periodicity of the quark fields up to the phase
$\exp(i\theta)$.  We thus have the angle $\theta$ as an additional
parameter that plays a r\^ole in the improvement condition for the
coefficient $\ca$\,\cite{paperIII}.

We are now prepared to define the expectation values of any
product $\cal O$ of fields by
\begin{equation}
  \langle{\cal O}\rangle=
  \left\{{1\over{\cal Z}}
  \int_{\rm fields}\,{\cal O}\,
  \rme^{-S}\right\}_
  {\rhobarprime=\rhoprime=\rhobar=\rho=0}.
\end{equation}
Apart from the gauge field and the quark and anti-quark fields
integrated over, $\cal O$ may involve the ``boundary fields" at time
$x_0=0$,
\begin{equation}
   \zeta({\bf x})={\delta\over\delta\rhobar({\bf x})},
  \qquad\kern3.5ex
  \zetabar({\bf x})=-{\delta\over\delta\rho({\bf x})},
\end{equation}
and similarly the fields at $x_0=T$. Note that the functional
derivatives only act on the Boltzmann factor, because the functional
measure is independent of the boundary values of the fields.

\subsection{Continuum limit and improvement 
 of the Schr\"odinger functional}
\myvspace
Based on the work of Symanzik~\cite{SchrodingerI,SchrodingerII} and
explicit calculations to one-loop order of perturbation
theory~\cite{alphaI,StefanI} one expects that the SF is renormalized
if the coupling constant and the quark masses are renormalized in the
usual way and the quark boundary fields are scaled with a
logarithmically divergent renormalization constant.

As in the case of the infinite volume theory discussed in
subsect.\,\ref{s_OSI}.2, the cutoff dependence of the SF may be described by a
local effective theory. An important difference is that the O($a$)
effective action $S_1$ now includes a few terms localized at the
space-time boundaries\,\cite{paperI}. Such terms then also appear in
the O($a$) improved lattice action. However, by an argument similar to
the one given at the end of subsect.\,\ref{s_OSI}.5, it can be shown that they
only contribute at order $a^2$ to the PCAC relation and the chiral Ward
identity considered in sect.\,5. In the calculations reported below,
the inclusion of boundary counterterms is, therefore, not required.

\section{IMPROVEMENT OF ACTION AND AXIAL CURRENT \label{s_IAAC}}

We now proceed to sketch the non-perturbative calculation of 
the coefficient $\csw$~\cite{paperI,paperIII}.

Using the operator
\begin{equation}
 {\cal O}^a
 =a^6\sum_{\bf y,z}\zetabar({\bf y})\dirac 5{{\tau^a}\over{2}}\zeta({\bf z}),
 \label{e_O}
\end{equation}
we define the bare correlation functions
\begin{equation}
 \begin{array}{l}
   \mbox{$\displaystyle\fa(x_0)\,\,=\,\,-\frac{1}{3}
   \langle A_0^a(x)\,{\cal O}^a\rangle,$}
 \\[1ex]
   \mbox{$\displaystyle\fp(x_0)\,\,=\,\,-\frac{1}{3}
   \langle P^a(x)\,{\cal O}^a\rangle.$}
 \end{array}
\end{equation}
In terms of $\fa$ and $\fp$ the PCAC relation for the unrenormalized
improved axial current and density may be written in the form
\begin{equation}
 m= {{\tilde{\partial}_0\fa(x_0)
 +\ca a\drvstar0\drv0\fp(x_0)}\over{2\fp(x_0)}}.
 \label{e_m}
\end{equation} 
We take this as the definition of the bare current quark
mass~$m$. The renormalized quark mass $\mr$ appearing in
eq.\,(\ref{e_PCAC_impr}) is then given by
\begin{equation}
  \mr=m{{\za(1+\ba a\mq)}\over{\zp(1+\bp a\mq)}}+\rmO(a^2).
\end{equation}
At fixed bare parameters, $\mr$ and hence also the unrenormalized
mass~$m$ should be independent of the kinematical parameters such as
$T,\,L$ and $x_0$. This will be true up to corrections of order $a^2$,
provided $\csw$ and $\ca$ have been assigned their proper values. The
coefficients may, therefore, be fixed by imposing the condition that
$m$ has exactly the same numerical value for three different choices
of the kinematical parameters.

For the rest of this section we set $T=2L$, $\theta=0$ and
$
  (\phi_1,\phi_2,\phi_3)=
  \frac{1}{6}\left(-\pi,0,\pi\right), \!
  (\phi'_1,\phi'_2,\phi'_3)=
  \frac{1}{6}\left(-5\pi,2\pi,3\pi\right)
  $.
An important practical criterion for choosing these particular 
boundary values has been that 
the induced background field should be weak on the scale of the lattice
cutoff to avoid large lattice effects.
On the other hand, the effects of order $a$, 
which one intends to cancel by adjusting $\csw$ 
should not be too small: otherwise one 
would be unable to compute $\csw$ accurately.
The above boundary values represent a compromise 
where both criteria are fulfilled
to a satisfactory degree on the accessible lattices.

Another mass, $\mprime$, 
may be defined in the same way by interchanging $C$ and $C'$.
The PCAC relation then implies that the mass difference
$\m-\mprime$ is of order $a^2$ if the coefficients $\csw$ and $\ca$
are appropriately chosen.
Our intention in the following is to take this as a condition 
to fix $\csw$.

%%%%%%%%%%%%%%%%%%%%%%%%%%%%%%%%%%%%%%%%%%%%%%%%%%%%%%%%%%%%%%%%%%%%%%
\begin{figure}[tb]
\vspace{-0pt}%\vspace{9pt}
\epsfig{file=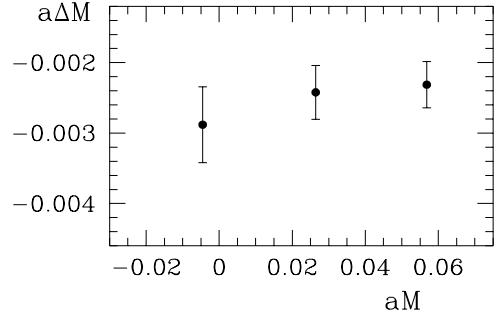,%18 144 592 718
       width=6.5cm}%
%      height=10cm}
%\framebox[55mm]{\rule[-21mm]{0mm}{43mm}}
\vspace{-30pt}
\caption{Mass difference $\dM$ 
on a $16\times8^3$ lattice as a function 
of the quark mass $\M$ at $\beta=6.4$ and $\csw=1.777$.
}
\label{f_deltam_m}
\end{figure}
%%%%%%%%%%%%%%%%%%%%%%%%%%%%%%%%%%%%%%%%%%%%%%%%%%%%%%%%%%%%%%%%%%%%%%

Before we are able to do so, we must eliminate the coefficient $\ca$,
which is not known either at this point. To this end 
first note that 
\begin{equation}
  \m(x_0)=\r(x_0)+\ca\s(x_0),
\end{equation}
where $r$ and $s$ are defined through
\begin{eqnarray}
  \r(x_0)&=&\frac{1}{4}(\drvstar{0}+\drv{0})\fa(x_0)/\fp(x_0),
  \\
  \s(x_0)&=&\frac{1}{2}a\drvstar{0}\drv{0}\fp(x_0)/\fp(x_0) \label{e_sx0}
\end{eqnarray}
(for clarity 
the dependence on the time coordinates is now often indicated explicitly).
The other mass $\mprime$ is similarly given in terms of two
ratios $\rprime$ and $\sprime$.
It is then trivial to show that the combination
\begin{equation}
  \M(x_0,y_0)=
  \m(x_0)-\s(x_0){\m(y_0)-\mprime(y_0)\over\s(y_0)-\sprime(y_0)}
\end{equation}
is independent of $\ca$, viz.
\begin{equation}
  \M(x_0,y_0)=
  \r(x_0)-\s(x_0){\r(y_0)-\rprime(y_0)\over\s(y_0)-\sprime(y_0)}.
\end{equation}

%%%%%%%%%%%%%%%%%%%%%%%%%%%%%%%%%%%%%%%%%%%%%%%%%%%%%%%%%%%%%%%%%%%%%%
\begin{figure}[tp]
\vspace{-00pt}%\vspace{9pt}
\epsfig{file=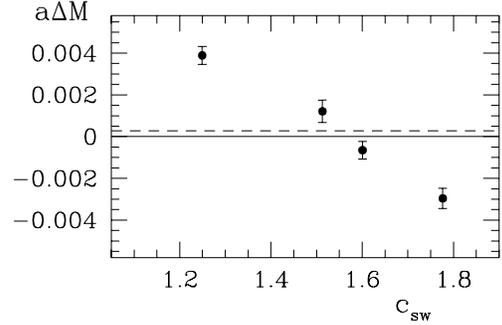,%18 144 592 718
       width=6.5cm}%
%      height=10cm}
%\framebox[55mm]{\rule[-21mm]{0mm}{43mm}}
\vspace{-30pt}
\caption{Determination of $\csw$ at $\beta=6.4$.
The dashed line indicates the tree-level value
$\dM^{(0)}$ appearing on the right-hand side of the improvement condition.
}
\label{f_deltam_csw}
\end{figure}
%%%%%%%%%%%%%%%%%%%%%%%%%%%%%%%%%%%%%%%%%%%%%%%%%%%%%%%%%%%%%%%%%%%%%%

Furthermore, from eq.~(\ref{e_sx0}) one infers that $\M$ coincides with $m$ up to a 
small correction of order $a^2$ (in the improved theory);
$\M$ may hence be taken as an 
alternative definition of an unrenormalized current 
quark mass, the advantage being that we do not need to know $\ca$
to be able to calculate it.
 
We now continue to discuss the condition that determines $\csw$. 
If we define $\Mprime$ in the same way as $\M$,
with the obvious replacements,
it follows from the above that the difference
\begin{equation}
  \dM=\M\big(\frac{3}{4}T,\frac{1}{4}T\big)-
      \Mprime\big(\frac{3}{4}T,\frac{1}{4}T\big)
\end{equation}
must vanish, up to corrections of order $a^2$, 
if $\csw$ has the proper value.
The coefficient may hence be fixed by demanding $\dM=\dM^{(0)}$.
Here $\dM^{(0)}$, the value of $\dM$ at tree-level of perturbation theory
in the $\Oa$ improved theory, is chosen instead of zero, in order to cancel a
small tree-level $\Oa$ effect in $\csw$. This is more a matter of aesthetics
than of practical importance (cf. fig.~\ref{f_deltam_csw}).
In order to complete the improvement condition,
we further give the precise definition of the
quark mass: we evaluate $\dM$ at
$\M\big(\frac{1}{2}T,\frac{1}{4}T\big)=0$. Again this choice is inessential,
changing $\csw$ only to $\rmO(aM)$, which are negligible effects
indeed (cf. fig.~\ref{f_deltam_m}).

At each of our nine values of $g_0$, we compute $\dM$ for $M=0$
for at least three values of $\csw$, and $\dM=\dM^{(0)}$
is solved for $\csw$ by a linear fit of $\dM$ as a function
of $\csw$. Representative data are displayed in
fig.\,\ref{f_deltam_csw}.

In the range $0\leq g_0^2\leq 1$, the results for $\csw$ are well
represented by\,\cite{paperIII}
\begin{equation}
 \csw={{1-0.656\,g_0^2-0.152\,g_0^4-0.054\,g_0^6}\over{1-0.922\,g_0^2}} .
 \label{e_csw_fit}
\end{equation}
%
%%%%%%%%%%%%%%%%%%%%%%%%%%%%%%%%%%%%%%%%%%%%%%%%%%%%%%%%%%%%%%%%%%%%%%
\begin{figure}[tp]
\vspace{-0pt}%\vspace{9pt}
\epsfig{file=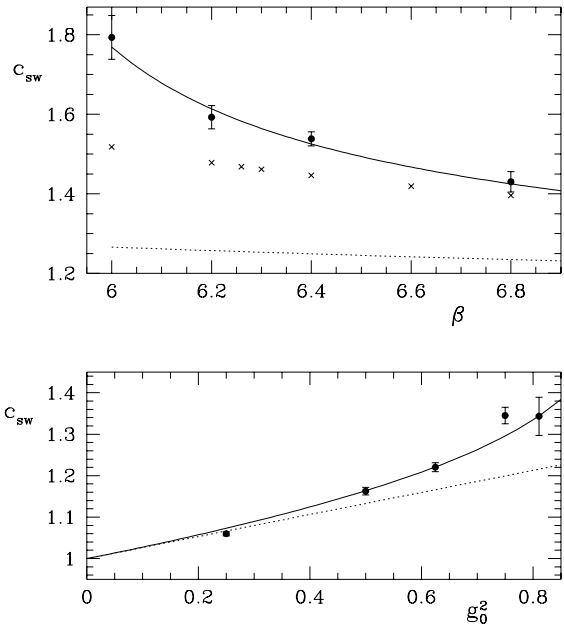,%18 144 592 718
       width=7.5cm}%
%      height=10cm}
%\framebox[55mm]{\rule[-21mm]{0mm}{43mm}}
\vspace{-30pt}
\caption{
  Non-perturbative improvement coefficient $c_{\rm sw}$.  
  Results from one-loop bare and tadpole improved perturbation theory
  are denoted by dotted lines and crosses, respectively.}
\label{f_csw}
\end{figure}
%%%%%%%%%%%%%%%%%%%%%%%%%%%%%%%%%%%%%%%%%%%%%%%%%%%%%%%%%%%%%%%%%%%%%%
%
In fig.\,\ref{f_csw} we compare our results to one-loop bare
perturbation theory and also to tadpole-improved perturbation
theory\,\cite{lepenzie_92}, for which we have used
\begin{equation}
   \csw = u_0^{-3}\left[ 1 + (\csw^{(1)}-1/4)\gparisi^2\right],
\end{equation}
where $\gparisi^2=g_0^2/u_0^4$\,\cite{Parisi}. Here $u_0^4$ is the
average plaquette in infinite volume.

\begin{figure}[tb]
\vspace{-0pt}%\vspace{9pt}
\epsfig{file=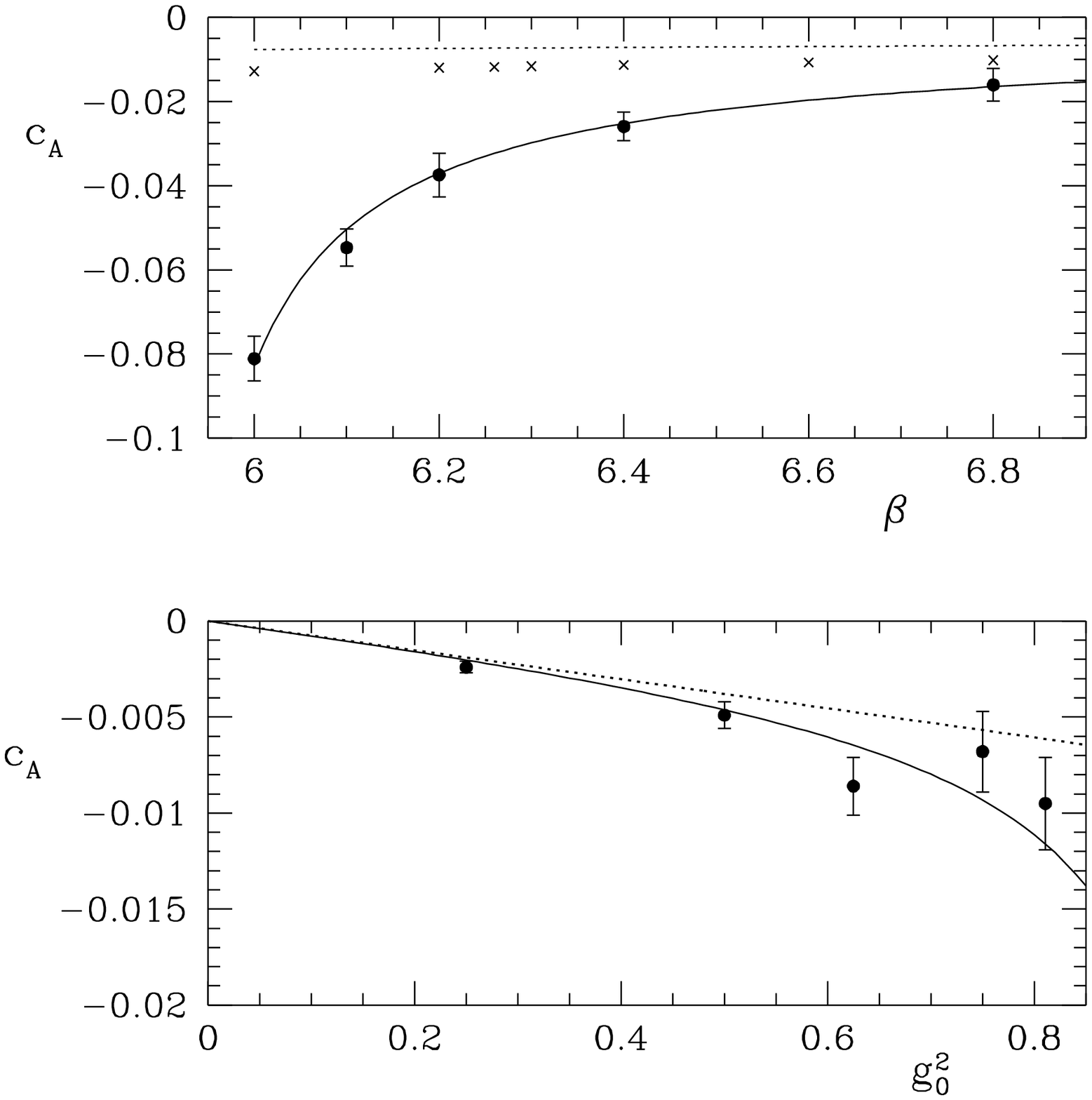,%18 144 592 718
       width=7.5cm}%
%      height=10cm}
%\framebox[55mm]{\rule[-21mm]{0mm}{43mm}}
\vspace{-30pt}
\caption{
  Non-perturbative improvement coefficient $\ca$.  
  Results from one-loop bare and tadpole improved perturbation theory
  are denoted by dotted lines and crosses, respectively.}
\label{f_ca}
\end{figure}

In a similar way\,\cite{paperIII} we obtained ($0\leq g_0^2\leq 1$)
\begin{equation}
 \ca=-0.00756\times g_0^2{{1-0.748\,g_0^2}\over{1-0.977\,g_0^2}},
 \label{e_ca_fit}
 \end{equation}
 as displayed in fig.~\ref{f_ca}.
Both eq.\,(\ref{e_csw_fit}) and eq.\,(\ref{e_ca_fit}) deviate 
substantially from the one-loop result except for values
of $g_0^2$ as small as $g_0^2\leq 1/2$.

%%%%%%%%%%%%%%%%%%%%%%%%%%%%%%%%%%%%%%%%%%%%%%%%%%%%%%%%%%%%%%%%%%%%%%
%%%%%%%%%%%%%%%%%%%%%%%%%%%%%%%%%%%%%%%%%%%%%%%%%%%%%%%%%%%%%%%%%%%%%%

\section{CURRENT NORMALIZATION}
\subsection{Chiral Ward identities}

For zero quark masses, chiral symmetry is expected to become exact in
the continuum limit.  It has therefore been proposed to fix the
normalization constants $\za$ and $\zv$ of the isospin currents by
imposing the continuum chiral Ward identities also at finite values of
the cutoff~\cite{BochicchioEtAl}--\cite{MartinelliEtAlI}.

In the case of the axial current the relevant Ward identity can be
written in the form
\begin{equation}
 \int_{\partial R}\!\!\!\!\rmd\sigma_\mu(x)\epsilon^{abc}
 \langle A_\mu^a(x)A_\nu^b(y) {\cal Q}^c\rangle 
  \!=\! 2i\langle V_\nu^c(y){\cal Q}^c\rangle,
\label{e_Ward}
\end{equation}
where the integral is taken over the boundary of the space-time region
$R$ containing the point $y$ and ${\cal Q}^a$ is a source located
outside $R$.  In view of on-shell improvement, it is important to note
that all space-time arguments in eq.\,(\ref{e_Ward}) are at non-zero
distance from one another.

For the source field in eq.\,(\ref{e_Ward}) we choose
${\cal Q}^a=\epsilon^{abc}{\cal O}'^{b}{\cal O}^{c}$
with ${\cal O}^a$ as given in eq.\,({\ref{e_O}), 
and ${\cal O}'{}^a$ defined similarly using the primed fields.
The region $R$ is taken to be $R=\{x,t_1 \leq x_0\leq t_2\}$.

We define the following correlation functions, using the
unrenormalized improved currents
$(A_{\rm I})_\mu^a=A_\mu^a+a\ca\tilde{\partial}_\mu P^a$ and 
$(V_{\rm I})_\mu^a=V_\mu^a+a\cv\tilde{\partial}_\nu T_{\mu\nu}$,
\begin{eqnarray}
  \fIaa(x_0,y_0)
         &=&-{{a^6}\over{6L^6}}\sum_{\bf x,y}\epsilon^{abc}\epsilon^{cde}
                    \nonumber\\
               & & \hspace{-0.3cm}\times\langle{\cal O}'^{d}
        (A_{\rm I})_0^a(x)(A_{\rm I})_0^b(y){\cal O}^{e}\rangle,
\end{eqnarray}
\vspace{-0.3cm}
\begin{equation}
  \fIv(x_0) = {{a^3}\over{6L^6}}\sum_{\bf x} i\epsilon^{abc}
                    \langle{\cal O}'^{a}(V_{\rm I})_0^b(x) 
                   {\cal O}^c\rangle,
\end{equation}
\vspace{-0.3cm}
\begin{equation}
  \f1       = -{{1}\over{3L^6}}\langle{\cal O}'^{a}{\cal O}^{a}\rangle.
\end{equation}

At $\mq=0$, and using the correct values of $\csw$ and $\ca$, 
a lattice version of the chiral Ward identity (\ref{e_Ward}) is
\begin{equation} \label{eq:za2}
 \za^2\fIaa(x_0,y_0)=\f1 +\rmO(a^2),\qquad x_0 > y_0 \, .
 \label{e_za}
\end{equation}
Compared to eq.\,(\ref{e_Ward}) we have set $\nu=0$ and included an
additional summation over ${\bf y}$, thus obtaining the isospin
charge. In deriving eq.\,(\ref{e_za}) we have used the fact that the
action of the latter on the chosen matrix elements can be evaluated
thanks to the exact isospin symmetry on the lattice. We further exploited the
conservation of the axial charge (at zero mass).

Since the isospin symmetry remains unbroken for non-zero quark mass,
one need not restrict the normalization condition of the vector
current to the case $\mq=0$. Using arguments similar to those in the
derivation of eq.\,(\ref{eq:za2}), one obtains
\begin{equation} \label{eq:zv}
 \zv(1+\bv a\mq)\fIv(x_0)=\f1 +\rmO(a^2).
\end{equation}
Note that the improvement coefficient $\cv$ is not needed here,
because the tensor density
does not contribute to the isospin charge.

\subsection{Normalization conditions}

Starting from eqs.\,(\ref{eq:za2},\,\ref{eq:zv}) we impose the
following normalization conditions on the axial and vector currents
at vanishing quark mass, $\mq=0$,
\begin{eqnarray}
 \za^2\fIaa(\frac23T,\frac13T) & = & \f1, \quad T=9L/4, \label{e_za1}\\
 \zv\fIv(\frac12T) & = & \f1, \quad T=2L, \label{e_zv1}
\end{eqnarray}
where we have set $C=C'=\theta=0$.  
In order to guarantee that cutoff effects of matrix elements
of the renormalized currents vanish proportional to $a^2$ when
approaching the continuum limit, we complete the normalization conditions
eqs.~(\ref{e_za1},\ref{e_zv1}) by scaling $L$ in units of a physical scale,
\begin{eqnarray}
L/a=8 \,\, {\rm at} \,\, g_0=1, \nonumber \\
L/r_0= \,{\rm constant}. \label{e_Lr0}
\end{eqnarray}
Here $r_0 \approx 0.5\,\fm$ is derived from the force between static quarks
as explained in ref.~\cite{r0}.
In the numerical simulations, the choices (\ref{e_za1}--\ref{e_Lr0})
together with a specific definition
of $\hopc$ are realized by smooth inter-/extrapolations~\cite{paperIV}.

Of course, the details of these choices are irrelevant.
Nevertheless, one must take care not to induce large $\rmO(a^2)$
effects through an ``unfortunate'' choice. For this reason the 
Ward identities were studied in perturbation theory before fixing the above
details. In addition we checked in the course of the simulations
that $\zv,\za$ do not change appreciably when parameters such as (\ref{e_Lr0})
or $\theta$ are changed within reasonable limits.

\subsection{Results}

Our numerical results are again well represented by
rational functions 
\begin{eqnarray}
  \zv&=&{1 - 0.7663 \,g_0^2 + 0.0488 \,g_0^4 \over 1 - 0.6369 \,g_0^2}, 
   \label{e_zvfit} \\
  \za&=&{1 - 0.8496 \,g_0^2 + 0.0610 \,g_0^4 \over 1 - 0.7332 \,g_0^2},
   \label{e_zafit}
\end{eqnarray}
which take into account the perturbative 
expansions~\cite{GabrielliEtAl,GoeckelerEtAl,StefanNotes}
\begin{eqnarray}
  \zv=1-0.129430 g_0^2+\rmO(g_0^4),
    \\
  \za=1-0.116458 g_0^2+\rmO(g_0^4).
\end{eqnarray}
Eq.~(\ref{e_zvfit}) and eq.~(\ref{e_zafit}) 
are to be quoted with total errors of 0.5\% and 1.0\%, respectively.

Fig.\,\ref{f_bv} shows the computed 
improvement coefficient $\bv$ \cite{paperIV} and the fit \cite{paperIV,paperV}
\begin{equation}
  \bv={{1-0.7613\,g_0^2+0.0012\,g_0^4-0.1136\,g_0^6}\over{1-0.9145\,g_0^2}},
\end{equation}
valid for $0\leq g_0 \leq1$.
%%%%%%%%%%%%%%%%%%%%%%%%%%%%%%%%%%%%%%%%%%%%%%%%%%%%%%%%%%%%%%%%%%%%%%
\begin{figure}[tb]
\vspace{-0pt}%\vspace{9pt}
\epsfig{file=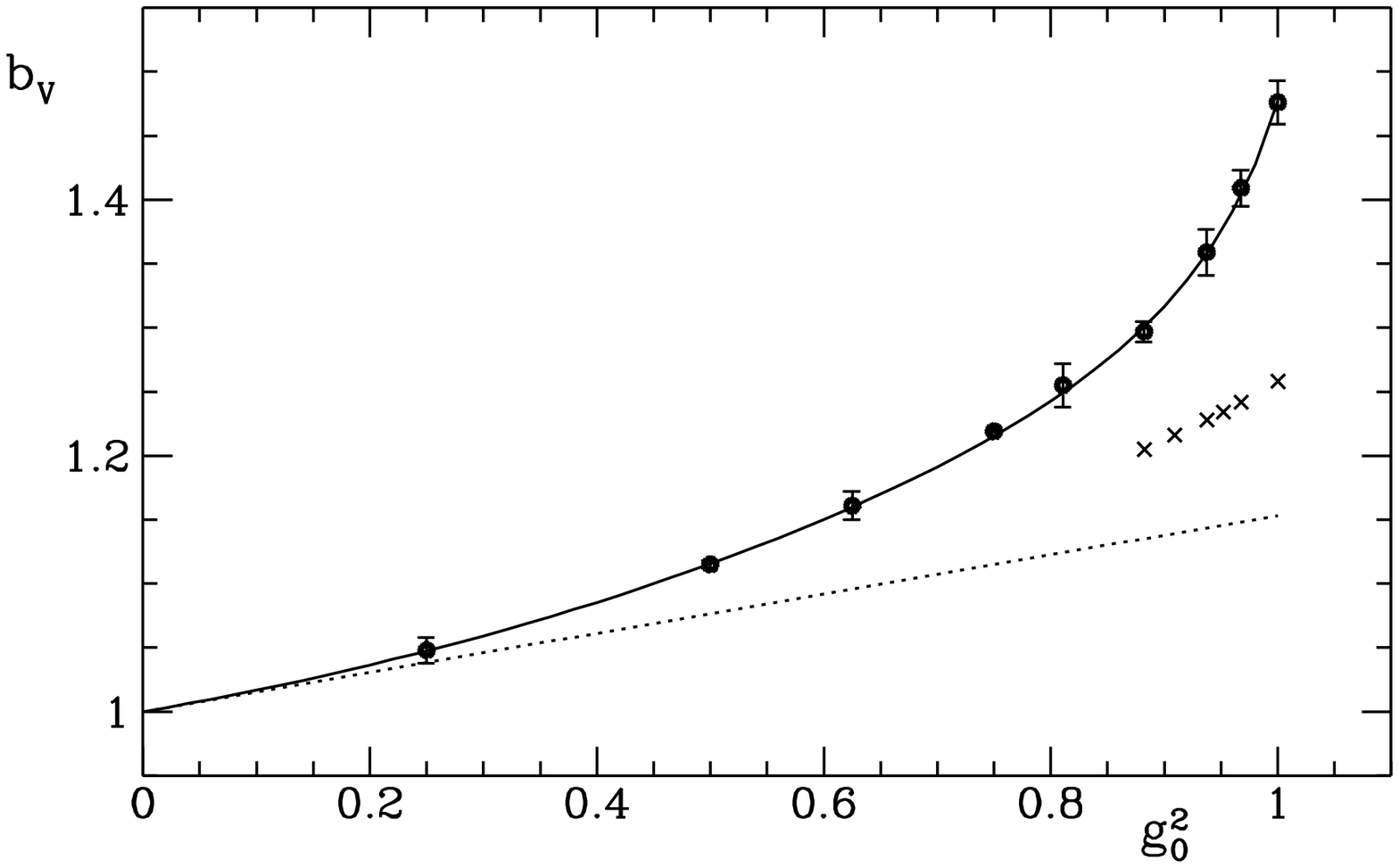,%18 144 592 718
       width=7.5cm}%
%      height=10cm}
%\framebox[55mm]{\rule[-21mm]{0mm}{43mm}}
\vspace{-30pt}
\caption{
  Non-perturbative improvement coefficient $\ca$.  
  Results from one-loop bare and tadpole improved perturbation theory
  are denoted by dotted lines and crosses, respectively.}
\label{f_bv}
\end{figure}
%%%%%%%%%%%%%%%%%%%%%%%%%%%%%%%%%%%%%%%%%%%%%%%%%%%%%%%%%%%%%%%%%%%%%%

Unfortunately, $\ba$ cannot be obtained in the same 
way,
since the relevant Ward
identity contains a physical mass dependence apart from the 
$\rmO(am)$ lattice artefact. We do not know how to separate the two. 

\section{IMPROVEMENT OF THE VECTOR CURRENT \label{s_IVC}}

For a complete determination of the renormalized improved 
vector current (\ref{e_renfields}), 
we further need the improvement coefficient $\cv$.
A computation of $\cv$ is currently in 
progress~\cite{cv}.
Here we outline the improvement condition employed and present
preliminary results.

The general idea is again quite simple. 
We have already seen above that correlation functions
of axial current and vector current
are related by chiral Ward identities. These are valid up to 
terms of order $a^2$ in the improved theory. 
Since the improved axial current is known, 
$\cv$ can be 
obtained through a suitable Ward identity. 

In order to excite states with vector quantum numbers, we
use the boundary operator
\begin{equation}
 ({\cal O}_{\rm V})^a_k
 =a^6\sum_{\bf y,z}\zetabar({\bf y})\dirac k{{\tau^a}\over{2}}\zeta({\bf z}),
 \label{e_Ok}
\end{equation}
and define the bare, unimproved, correlation functions 
\begin{eqnarray}
 \kv(x_0)&=&- {1\over9} \langle V_k^a(x)\,({\cal O}_{\rm V})^a_k\rangle , \\
 \kt(x_0)&=&- {1\over9} \langle T_{k0}^a(x)\,({\cal O}_{\rm V})^a_k\rangle 
\end{eqnarray}
as well as the bare, improved correlation function 
\begin{eqnarray}
 \kaaI(x_0,y_0)&=&{i\over18} a^3 \sum_{\bf x}\epsilon^{abc}  \\
 && \times \langle (\ai)_0^a(x) (\ai)_k^b(y) 
                 \,({\cal O}_{\rm V})^c_k\rangle.\nonumber
\end{eqnarray}
%%%%%%%%%%%%%%%%%%%%%%%%%%%%%%%%%%%%%%%%%%%%%%%%%%%%%%%%%%%%%%%%%%%%%%
\begin{figure}[tb]
\vspace{-0pt}%\vspace{9pt}
\epsfig{file=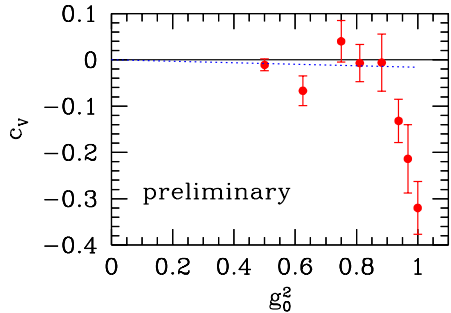,%18 144 592 718
       width=6.0cm}%
%      height=10cm}
%\framebox[55mm]{\rule[-21mm]{0mm}{43mm}}
\vspace{-30pt}
\caption{
  Non-perturbative improvement coefficient $\cv$.  
  One-loop bare perturbation theory is 
  shown as a dotted line.}
\label{f_cv}
\end{figure}
%%%%%%%%%%%%%%%%%%%%%%%%%%%%%%%%%%%%%%%%%%%%%%%%%%%%%%%%%%%%%%%%%%%%%%
These correlation functions are related through the Ward identity
($t_1 < x_0 < t_2$)
\begin{eqnarray}
 &\zv& \left[\kv(x_0)+a \cv \tilde{\partial}_0 \kt(x_0)\right]  
  \label{e_WIcv}\\
  &&\, = \za^2 \left[\kaaI(t_2,x_0) - \kaaI(t_1,x_0)\right] + \rmO(a^2)
  ,\nonumber
\end{eqnarray}
which is valid in the massless limit.  Eq.~(\ref{e_WIcv}) is derived in 
much the same way
as the Ward identity discussed in the
previous section. Inserting $\za,\zv$ and $\ca$, it can be solved for 
the improvement coefficient $\cv$.

The reader will by now be aware
that, for a precise definition of the improvement condition, we
have to make specific choices for $L,T,\theta,\ldots$.
These, as well as a number of other 
details, will be given in ref.~\cite{cv}.

First results (cf. fig.~\ref{f_cv}) show $\cv$ to
be much larger than the perturbative 
estimate eq.~(\ref{e_cApert}) for couplings $g_0^2>0.9$.
Its effect on vector meson decay constants and semileptonic decay 
form factors may be significant in simulations with presently
achievable resolutions.

\section{$\rmO(a^2)$  EFFECTS}

In sect.~2 we gave a perturbative example, where 
$\Oa$ improvement removed most of the cutoff effects, leaving only
tiny effects of higher order in the lattice spacing.  Of course, it is 
important to check in how far this is true non-perturbatively. 

\subsection{The PCAC relation}

The first test of $\rmO(a^2)$ effects is provided again
by the PCAC relation. To set the scale, remember that the cutoff
effects in the PCAC mass $m$ were as large as several tens of $\MeV$
before improvement (cf. fig.~\ref{f_letter}). The situation 
after improvement, and for a somewhat larger value of the lattice spacing,
is illustrated in fig.~\ref{f_mimpr}. Away from the boundaries, where
the effect of states with energies of the order of the cutoff induces
noticeable effects, $m$ is independent of time to within $\pm 2\,\MeV$.

%%%%%%%%%%%%%%%%%%%%%%%%%%%%%%%%%%%%%%%%%%%%%%%%%%%%%%%%%%%%%%%%%%%%%%
\begin{figure}[tb]
\vspace{-0pt}%\vspace{9pt}
\epsfig{file=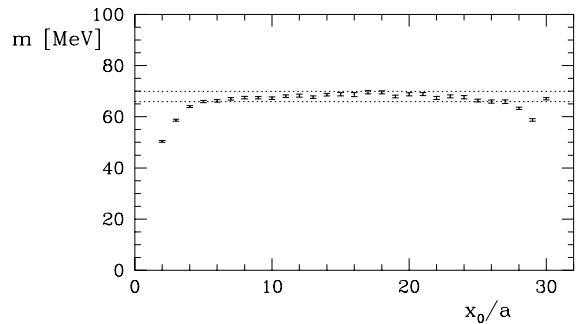,%18 144 592 718
       width=7.5cm}%
%      height=10cm}
%\framebox[55mm]{\rule[-21mm]{0mm}{43mm}}
\vspace{-30pt}
\caption{
Unrenormalized current quark mass $\m$ 
in the improved theory, with non-perturbatively determined $\csw$
and $\ca$, as a function 
of the time $x_0$ on a $32\times16^3$ lattice
at $\beta=6.2$ and $\hop=0.1350$. The width of the corridor bounded by the 
dotted horizontal lines is $4\,\MeV$.
}
\label{f_mimpr}
\end{figure}
%%%%%%%%%%%%%%%%%%%%%%%%%%%%%%%%%%%%%%%%%%%%%%%%%%%%%%%%%%%%%%%%%%%%%%

A search for other lattice artefacts after improvement
yielded, as largest effect, the one
described in the following.

%%%%%%%%%%%%%%%%%%%%%%%%%%%%%%%%%%%%%%%%%%%%%%%%%%%%%%%%%%%%%%%%%%%%%%
\begin{figure}[tb]
\vspace{-0pt}%\vspace{9pt}
\epsfig{file=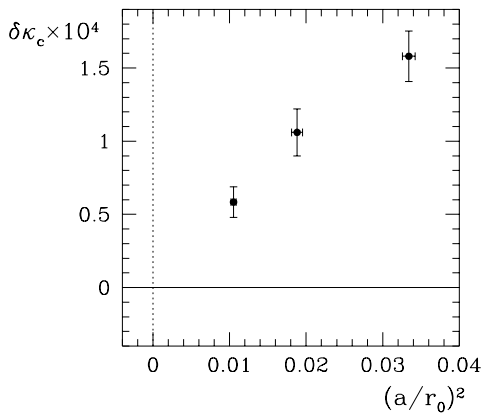,%18 144 592 718
       width=6.0cm}%
%      height=10cm}
%\framebox[55mm]{\rule[-21mm]{0mm}{43mm}}
\vspace{-30pt}
\caption{
Change $\delta\hopc$ in the calculated value of $\hopc$ when
$L/a$ is in\-creased from $8$ to $16$. The data have been obtained 
at $\beta=6.0,6.2$ and $6.4$ (from right to left).
For $\beta>6.8$, $\delta\hopc$ is consistent with zero within the statistical
errors that are at most $10^{-5}$.
}
\label{f_dkc}
\end{figure}
%%%%%%%%%%%%%%%%%%%%%%%%%%%%%%%%%%%%%%%%%%%%%%%%%%%%%%%%%%%%%%%%%%%%%%
As chiral symmetry is violated by lattice artefacts, there is no
precise definition of a chiral point
$\kappa=\kc\equiv1/(8+2a\mc)$ for any finite value of the lattice
spacing\,\cite{paperI}.  Rather, $\kc$ has an intrinsic
uncertainty, which is reduced from $\rmO(a^2)$ to $\rmO(a^3)$ by
non-perturbative improvement.

Here, we define $\kc$ as the value of $\kappa$, where $m$,
eq.\,(\ref{e_m}), vanishes for $C=C'=\theta=0, T=2L, x_0=T/2$.  To
study the $\rmO(a^3)$ effects we can then still vary the resolution
$a/L$.

Close to $\kc$ the mass $m$ is a linear function of $\kappa$ to a high
degree of accuracy. The point $\kc$ is therefore easily found by
linear interpolation (or, for $\beta \leq 6.2$, slight extrapolation).

For $\beta>6.8$, the results for $a/L=1/8$ and
$a/L=1/16$ agree on the level of their statistical accuracy, which is
better than $10^{-5}$. For lower $\beta$, small but significant
$\rmO(a^3)$ effects are seen (fig.\,\ref{f_dkc}). 
They decrease rapidly as the lattice spacing
is reduced.\footnote{ 
Note that the dependence of $\delta\hopc$ on $a/r_0$ 
is non-trivial since there are two relevant physical scales, $L$ and $r_0$,
in the regime covered by fig.~\ref{f_dkc}.
On general grounds
one can only predict that $\delta\hopc \sim (a / r_0)^3 d(L/r_0)$ with some
unknown function $d$. }

\subsection{Hadronic observables}

It is of great interest to check the size of
$\rmO(a^2)$ terms in hadronic observables. At this conference first results 
for the improved theory have
been presented by UKQCD~\cite{Richard} and we had the pleasure to survey
the calculations of De~Divitiis et al.~\cite{roma2} and G\"ockeler et 
al.~\cite{qcdsf}.
Since all of these results are still preliminary, we restrict ourselves 
to some
qualitative observations. 

Wherever a comparison is possible,
the results of the different groups are in good 
agreement.
As the 
calculations were initiated only fairly recently, results to 
date exist for 
two values of the lattice spacing only, which
correspond to
$(a/r_0)^2=0.035$ and $(a/r_0)^2=0.019$. 
Although the lattice artefacts should therefore vary by roughly a factor
of 2 between the two, a third, smaller value of $a$ appears necessary
to perform  precise
extrapolations to the continuum limit. 

Nevertheless, one observes~\cite{Richard,qcdsf} 
that the lattice spacing dependence
of the ratio of the vector meson mass to the
square root of the string tension 
is reduced by improvement; indeed the values calculated
in the improved theory are close to 
those computed without improvement and extrapolated to $a=0$.
Unfortunately, it is not obvious that systematic errors 
in the determination of the 
string tension do not distort the picture. It would be desirable to repeat the
analysis, replacing the string tension by $r_0$~\cite{r0}.  

The effect of the improvement of the axial current has been studied
in the calculation of $f_{\pi}$. Despite the rather small value of the 
coefficient $\ca$, the admixture of $\partial_\mu P$ represents a 10\% 
change in $f_\pi$
at $(a/r_0)^2=0.035$. After inclusion of this effect, 
$r_0 f_\pi$  is independent of the  lattice spacing 
within the statistical errors of around 3\%.
 
In summary no large lattice artefacts have  so far been found in
the (quenched) improved theory. Improvement (and of course proper
renormalization) of the composite fields is important in this respect.  

\section{DYNAMICAL FERMIONS \label{s_nf2}}

The ALPHA collaboration has started to implement
improvement in full QCD with two flavours. 

In particular, we are currently calculating $\csw(g_0)$ along the lines 
described 
in section~\ref{s_IAAC}~\cite{csw_dyn}. 
We simulate with the 
Hybrid Monte Carlo algorithm~\cite{HMC} employing even-odd preconditioning 
of the Dirac operator
and the Sexton--Weingarten scheme~\cite{SextWein}
for the integration of the equations of motion. 
The classical trajectories have 
unit length. For details of the
implementation see ref.~\cite{HMC_impl}. Alternative algorithms are being
studied at the same time~\cite{KarlRoberto}.

One slight difference from sect.~\ref{s_IAAC} is that we do not 
insist on setting the improvement condition at $M=0$,
but rather keep $aM$ small, say smaller than 0.02. Remember
that the dependence of $\Delta M$ on $M$ is insignificant (cf.
fig.~\ref{f_deltam_m}).  
The reason for not attempting to put $M$ to zero exactly
is {\it not} a difficulty in simulating in the massless limit.
In the SF the massless Dirac operator 
has a gap~\cite{StefanI}, at least at not too large values of $L$,
say $L\leq 1\,\fm$. Indeed, the simulations done so far
show that it is irrelevant for the simulations whether we 
have $aM=0.02$ or $aM=0.001$.
We do choose finite values of $M$ in order to avoid an
unnecessarily large effort for
tuning $M$.

Having said this, we mention, however, that as $g_0$ is increased at fixed
$L/a=8$, the physical scale
$L$ increases and the gap in the Dirac operator
becomes smaller. This means that more
Conjugate Gradient iterations are needed to obtain a solution
of the Dirac equation. Quantitatively, in one hour on a 256-node APE-100, 
we obtain around 100 trajectories at $g_0^2=0.8$, 50 trajectories at 
$g_0^2=1$ and ``only'' 30 trajectories at $g_0^2=1.1$.

At the same time our autocorrelation times for typical long-range
observables grow from roughly 1 trajectory at $g_0^2=0.5$
to 4 trajectories at $g_0^2=1.05$.  

The simulations were
started at small values of $g_0$ and reach
$g_0^2=1.1$ by now.

In 1-loop perturbation theory~\cite{StefanRainer}, there is
a shift in the bare coupling $\beta=6/g_0^2$ from the Wilson action
to the improved action of $-0.3$ ($\nf=2$). Using this
and estimates of the lattice spacing with Wilson fermions~\cite{SESAM}, 
a rough
estimate for the lattice spacing at $g_0^2=1.1$ is $0.1\,\fm$.
We emphasize that this is a rough estimate since the true
renormalization 
may be quite different at such a large value of the coupling. 
Nevertheless this indicates that $\csw$ will not be needed for
values of $g_0^2$ that are much larger than $g_0^2=1.1$.

\section{CONCLUSIONS}

We have been able to implement on-shell O($a$) improvement
non-perturbatively in (quenched) lattice QCD. The improvement
coefficients $\csw$, $\ca$, the critical mass $\mc$ and the current
normalization constants $\za$ and $\zv$ have been determined for bare
couplings in the range $0\leq g_0\leq1$.  In all cases, contact with
bare perturbation theory could be made at couplings around $g_0^2
\approx 0.25 - 0.5$.  The convergence of perturbation theory appears
to be significantly speeded up by tadpole improvement as shown by the
crosses in figs.~\ref{f_csw}--\ref{f_bv}. However, for values of
$\beta \le 6.8$, which is the range most relevant for present MC
computations, the quality of tadpole-improved perturbation theory is
rather non-universal. There are examples such 
as $\bv, \ca$, where  perturbation
theory is far off from our results. Renormalizations are large!

This suggests that one should be cautious when estimating
improvement terms to a low order of perturbation theory, as
is currently done in several attempts to extend improvement to the  
$\rmO(a^2)$ level.

It is very pleasing that non-perturbative results can be obtained with 
statistical errors that are much smaller than the intrinsic 
$\rmO(\alpha^2)$ uncertainties of the perturbative results for the improvement
coefficients. 
The elimination of these perturbative uncertainties allows
for reliable continuum extrapolations of the hadron spectrum and 
current matrix elements to be done in the future. 
Indeed, as a first step in this direction
it was observed 
that residual $\rmO(a^2)$ effects in hadronic quantities 
appear to be quite small. 
In particular, we point out that the $a$-independence of $r_0f_{\pi}$ was
observed after accounting for the relatively large correction 
proportional to $\ca$, a term that would have been set to zero in 
tree-level improvement and is still tiny at 1 loop.   
\\[1.5ex]    
Acknowledgements\\[0.5ex]
I thank the organizers of this workshop for creating a very stimulating 
atmosphere and all the other members of the
ALPHA collaboration for sharing their knowledge with me.
I am grateful to M. Massetti and G. Schierholz for sending 
me results prior to publication.
Most of the calculations discussed here have been obtained in the course of
the ALPHA collaboration research programme. We
thank DESY for allocating computer time on the APE-Quadrics computers
to this project, and are grateful to the staff of the computer centre
at DESY-IfH, Zeuthen for their support.

\end{document}